\documentstyle[11pt,twoside,psfigack,jltp]{article}

\runninghead{J.M. Elzerman \boldmath$et$ $al$.}{Suppression of the Kondo Effect by Microwave Radiation}
\begin{document}

\title{Suppression of the Kondo Effect in a Quantum Dot by Microwave Radiation}
\author{Jeroen M. Elzerman, Silvano De Franceschi,
\newline 
David Goldhaber-Gordon \footnote{
Present address: Department of Physics, Harvard University,                                  
17 Oxford Street, Cambridge MA 02138.                                
},
Wilfred G. van der Wiel, 
\newline
and Leo P. Kouwenhoven%
\address{Department of Applied Physics and DIMES, Delft
University of Technology, 
\newline
PO Box 5046, 2600 GA Delft, The Netherlands}}

\begin{abstract}
We have studied the influence of microwave radiation on the transport
properties of a semiconductor quantum dot in the Kondo regime. In the entire
frequency range tested (10---50 GHz), the Kondo resonance
vanishes by increasing the microwave power.
This suppression of the Kondo resonance shows an unexpected scaling behavior. 
No evidence for photon-sideband formation is found. The comparison with 
temperature-dependence measurements
indicates that radiation-induced spin dephasing plays an important role in
the suppression of the Kondo effect.

\end{abstract}

\maketitle

\section{INTRODUCTION}

The Kondo effect stems from a macroscopic quantum coherent coupling between
a localized magnetic moment and a Fermi sea of electrons. The local magnetic
moment is screened by hybridization with the delocalized electron-spins,
leading to the formation of a bound spin-singlet state. A many-body
resonance in the density of states (DOS) appears at the Fermi energy, which
strongly influences the conductance of the system at low temperatures. Above
the Kondo temperature, thermal fluctuations destroy the coherence, resulting
in a non-monotonic dependence of the conductance on temperature.

The equilibrium Kondo effect has been studied extensively in metals
containing a dilute concentration of magnetic impurities\cite{hewson}. In
addition, the dependence on bias voltage and magnetic field was studied in
single magnetic impurities encapsulated in a tunnel barrier\cite{rowell}. In
recent years, also semiconductor quantum dot devices have been considered,%
\cite{dgg}$^{-}$\cite{simmel} thanks to their high controllability. For
instance, a gate voltage can be used to change the number of confined
electrons, and hence switch the total spin of the dot between zero (even
electron number) and one half (odd electron number). In this way the Kondo
temperature can be tuned from arbitrarily small values to typically a few
Kelvin.

In this paper we investigate the Kondo effect in a quantum dot under
controlled microwave irradiation. This is motivated by several theoretical
studies,\cite{hettler}$^{-}$\cite{kaminski} which, interestingly, make
contradicting predictions. Assuming that radiation of frequency $f$ does not
lead to ionization of the dot and subsequent spin-decoherence, it is
expected that sidebands will appear in the DOS, at multiples of $hf$
away from the main Kondo resonance. These sidebands should be measurable as
additional peaks in the differential conductance.

In contrast to these predictions, Kaminski, Nazarov and Glazman\cite
{kaminski} have pointed out that even without ionization, 
absorption of radiation during a cotunneling process 
effectively flips the spin on the dot. This spin flip cuts off
the sequence of spin-correlated tunneling processes that build up the Kondo
resonance, thus giving rise to a finite lifetime of the spin-singlet state.
Consequently the Kondo effect is suppressed.

Now the question becomes whether in an experiment where the microwave power
is progressively increased, the sidebands are observable before the Kondo
effect is destroyed completely. Incidentally, low-power external radiation of unclear origin 
has been pointed out as the dominant cause for dephasing in weak
localization experiments at low temperatures\cite{altshuler}. 
One of these experiments was recently perfomed with open quantum dots under intentional microwave  
irradiation \cite{huibers}.
Photon sidebands have been observed in quantum dots weakly coupled to
electron reservoirs\cite{tjerk}. It should be stressed that in these
experiments tunneling is a very weak perturbation and coherence does not
play a role. In the present paper we employ the techniques outlined in Ref.
16 to investigate the still unexplored effects of microwaves in the
strong-tunneling Kondo regime.
\newpage
\section{THEORY}

\subsection{{\bf DC Kondo Effect in Quantum Dots}}

In a quantum dot with an odd number of electrons, the total spin is
necessarily non-zero, resulting in a magnetic moment localized on the dot.
This moment can be screened when exchange of electrons is possible with a
nearby reservoir, as shown schematically in Fig. 1a. The dot is simplified
to just one energy level, $\varepsilon _{0}$, which in Fig. 1a is occupied
by an electron with spin-up. Adding an electron increases the energy by the
on-site Coulomb interaction, $U$. The dot is coupled to two reservoirs with
electrochemical potentials $\mu _{L}$ and $\mu _{R}$, via tunnel barriers
characterized by the tunneling rates $\Gamma _{L}$ and $\Gamma _{R}$.

In the situation of Fig. 1a first-order tunneling is blocked by Coulomb
charging effects\cite{curacao}. Nevertheless, an electron can tunnel off the
dot through an intermediate virtual state, which costs an energy $\sim U$
and is allowed only for a short time $\sim h/U$. The system returns to a
low-energy state when an electron tunnels onto the dot. This can result in
an effective spin-flip. When many of such spin-flip tunneling events via
virtual states take place in a coherent way, the total spin state of the dot
plus reservoirs becomes a singlet, i.e. the local moment is completely
screened. This, in brief, is the Kondo effect\cite{hewson}.

The charge fluctuations associated with these higher-order tunneling events
lead to a broadening of the single-particle state $\varepsilon _{0}$ by an
amount $h\Gamma =h(\Gamma _{L}+\Gamma _{R})$. In addition, spin fluctuations
give rise to a narrow many-body resonance in the tunneling density of
states, located at the Fermi energies of the reservoirs (see Fig. 1b). 
The width of this resonance gives the  
typical energy scale for the Kondo effect,  
$\sim$$k_{B}T_{K}$, where $T_{K}$ is the Kondo temperature. In terms of dot parameters, $%
k_{B} T_{K}\approx \sqrt{h \Gamma U}\,e^{\pi (\varepsilon_{0}-\mu)
(\varepsilon_{0} + U - \mu)/2 h \Gamma U}$,
with $\mu =\mu _{L}=\mu _{R}$\cite{haldane}. At temperatures above $T_{K}$, the Kondo
effect is destroyed by thermal fluctuations.

In the linear regime, the Kondo resonance results in an enhanced
conductance, since the two reservoirs are always connected by a dot with a
finite DOS at the Fermi energy. Applying a bias voltage, $V_{sd}$, leads to
a finite lifetime of the spin-singlet state and thus to a suppression of the
Kondo effect. Experimentally, this translates into a peak in the
differential conductance, $dI/dV_{sd}$, centered around $V_{sd}=0$.

\subsection{\bf AC Kondo Effect in Quantum Dots}

Photon-assisted tunneling experiments on quantum dots\cite{tjerk} have shown
that microwave radiation can give rise to the formation of photon sidebands
(see Fig. 1c). Instead of just one single-particle state at $\varepsilon
_{0} $, additional states are formed at $\varepsilon _{0}\pm nhf$, where $n$
is an integer referring to the number of absorbed or emitted microwave
photons. The occupation of these sidebands depends on the intensity of the
radiation. We will only consider low powers, such that $n$ is restricted to $%
0$ or $\pm 1$.

Several theoretical papers\cite{hettler}$^{-}$\cite{kaminski} have addressed
the question what would happen if such experiments are repeated in the Kondo
regime. 
\begin{figure}[p]
\centerline{\psfig{figure=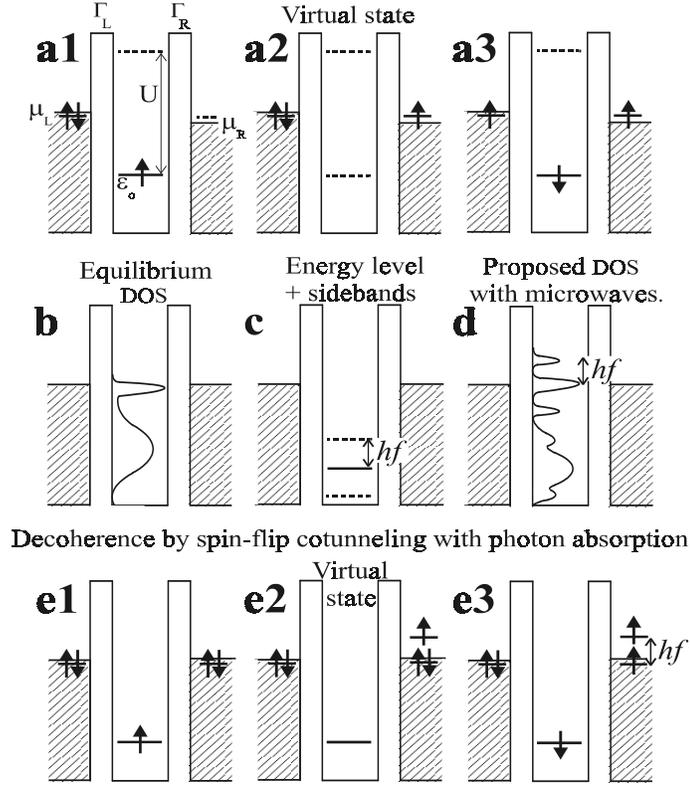,width=15cm,height=25cm,rheight=11.5cm}}
\caption{  
(a) Energy diagrams with one
spin-degenerate energy level $\varepsilon _{0}$ occupied by a single
electron. The series (a1, a2, a3) depicts an example of a higher-order
tunneling event, in which the spin-up electron tunnels off the dot through a
virtual intermediate state (a2), and a spin-down electron tunnels onto the
dot. All such events involving spin-flips contribute to a macroscopically
correlated state, the Kondo resonance. The resulting DOS is schematically
shown in (b). The broad lower bump is due to $\varepsilon _{0}$, whereas the
narrow peak at the Fermi energy of the leads represents the Kondo resonance.
(c) In the presence of microwaves of frequency $f$, sidebands of the level $%
\varepsilon _{0}$ develop at energies $\varepsilon _{0}\pm hf$. (d) Besides
these single-particle sidebands (the broad lower bumps in the DOS),
microwaves may also lead to many-body satellites of the Kondo resonance (the
narrow peaks at $\mu \pm hf$). (e) Spin-flip cotunneling process involving
the absorption of a photon. Such processes lead to spin decoherence.
}
\end{figure}
The answer is not immediately clear since the Kondo effect arises
from a spin-correlation between many subsequent tunneling events. 
The
question arises whether or not photon-assisted tunneling breaks this
coherence. Note that the formation of sidebands in the single-particle
spectrum makes this problem resemble 
the multi-level Kondo problem.\cite{inoshita}$^{,}$\cite{teemu} 
If the spin-coherence is not affected, the
sidebands lead to narrow satellites of the Kondo resonance, at energies $\mu
\pm hf$ (see Fig. 1d). These satellites should be observable in the $%
dI/dV_{sd}$ characteristics as additional peaks at $V_{sd}=\pm hf/e$.

However, the radiation can also induce decoherence in the spin-state of the
system. One mechanism is related to ionization of the dot; an electron can
be excited from the dot to the reservoirs by absorbing a photon, if the
frequency is high enough (i.e. $hf>\mu -\varepsilon _{0}$). A different
electron can then tunnel back onto the dot, but it has no spin-correlation
with the previous electron. This gives rise to a finite lifetime of the
spin-singlet state, leading to a suppression of the Kondo effect\cite
{nordlander}. Experimentally, this problem can quite easily be circumvented
by choosing $f$ low enough such that ionization is not allowed (i.e. $hf<\mu
-\varepsilon _{0}$).

Another, more subtle, mechanism for spin-decoherence is spin-flip cotunneling%
\cite{kaminski}, which is illustrated in Fig. 1e. It again involves the
absorption of a photon and a subsequent spin-flip, but the difference is
that now the intermediate state is virtual. As a result, this process is
possible regardless of the microwave frequency. It is therefore the dominant
decoherence-mechanism due to radiation at low frequencies.

It is important to note that the lifetime of the satellite peaks in the
differential conductance can be substantially smaller than that of the main
Kondo peak. This results from the fact that the satellites are measured at
finite bias, which allows for extra decoherence processes. The strength of
the satellite peaks depends in a complicated way on various parameters, such
as the microwave power, frequency and temperature. The question whether the
satellites of the main Kondo resonance are observable in experiments is
difficult to answer a priori.

\section{SAMPLE AND SETUP}

An image of the device is shown in Fig. 2. Six metallic gates are fabricated
on a GaAs/AlGaAs heterostructure containing a 2D electron gas (2DEG) about
100 nm below the surface. The electron density is $1.9\times 10^{15}$ m$%
^{-2} $. Appropriate negative gate voltages define a small puddle of
electrons, which is coupled to the 2DEG reservoirs via tunable tunnel
barriers. The lithographic size of the dot is $\sim $300$\times $300 nm$^{2}$%
. Taking into account a depletion region, we estimate an effective size of
about 170$\times $170 nm$^{2}$ with approximately 60 electrons in the dot.

\begin{figure}[t]
\centerline{\psfig{figure=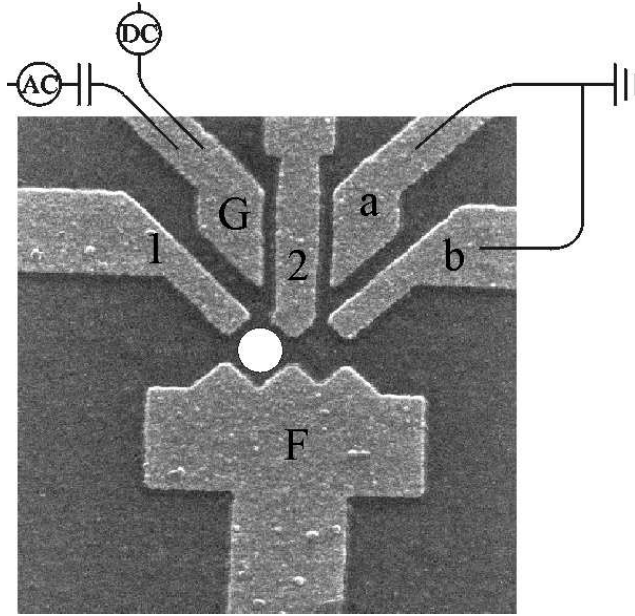,width=12cm,rheight=9.0cm}}
\caption{  
SEM image
of the quantum dot device. Light gray regions indicate the metallic gates,
dark gray regions the surface of the GaAs/AlGaAs heterostructure. The white
circle indicates the electron puddle. Both a DC and an AC voltage can be
applied to the plunger gate G.}
\end{figure}
The two gates labelled {\it a} and {\it b} are grounded in the present
experiments. Gates 1, 2 and {\it F} control the size of the dot and the
tunnel barriers. A DC voltage, $V_{g}$, applied to the plunger gate, {\it G,}
is used to finely tune the electron number. In addition, an AC voltage of
amplitude $V_{\omega }$ and frequency up to 50 GHz can be applied by
coupling a high-frequency coaxial cable capacitively to the plunger gate.
The AC power is controlled with a variable attenuator. Additional fixed
attenuators are inserted in the coaxial line at low temperatures. The effect
of the microwaves is measured in DC transport. A current, $I,$ is measured
in response to a voltage, $V_{sd}$, applied between the source and drain
reservoirs, using a standard lock-in technique with 3 $\mu $V excitation at
17.7 Hz. The device was mounted in a dilution refrigerator, and temperature
was varied between 18 mK and 600 mK. We have applied a magnetic field of
0.15 T, to lower the resistances of the Ohmic contacts. This field is small
enough not to affect the Kondo resonance, i.e. the Zeeman splitting $g\mu
_{B}B\ll k_{B}T_{K}$.

\begin{figure}[t]
\centerline{\psfig{figure=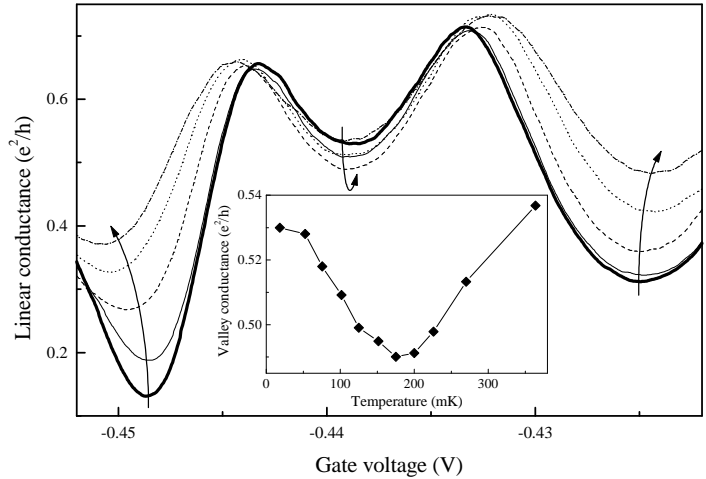,width=16cm,rheight=6.7cm}}
\caption{  
Linear conductance, $I/V_{sd}$,
versus gate voltage, $V_{g}$, for various temperatures: $T$ (in mK) = 18
(thick), 102 (thin), 175 (dashed), 270 (dotted), 364 (dashed-dotted). Each
curve has been given an offset in the x-direction to compensate for a
temperature-dependent shift\protect\cite{T-shift}. Arrows indicate the
direction of increasing $T$. {\it Inset}: Conductance in the middle of the
central valley versus temperature.}
\end{figure}

\section{MEASUREMENTS}

\subsection{DC Kondo Effect}

We make the tunnel barriers rather transparent, to increase the 
tunnel rate $\Gamma $ and thus $%
T_{K}$. This leads to broad and overlapping Coulomb peaks, as in Fig. 3,
which shows the linear conductance versus gate voltage for various
temperatures between base temperature, $18$ mK, and $365$ mK \cite{T-shift}.

At base temperature (thick line), the two Coulomb peaks are paired. When the
temperature, $T$, is raised, the conductance in the outer valleys increases
(indicated by arrows). In the central valley, on the other hand, the
conductance first decreases until it starts to increase for $T > 200$ mK.
Also, the Coulomb peaks move apart. These are clear signs of the Kondo
effect.

\begin{figure}[t]
\centerline{\psfig{figure=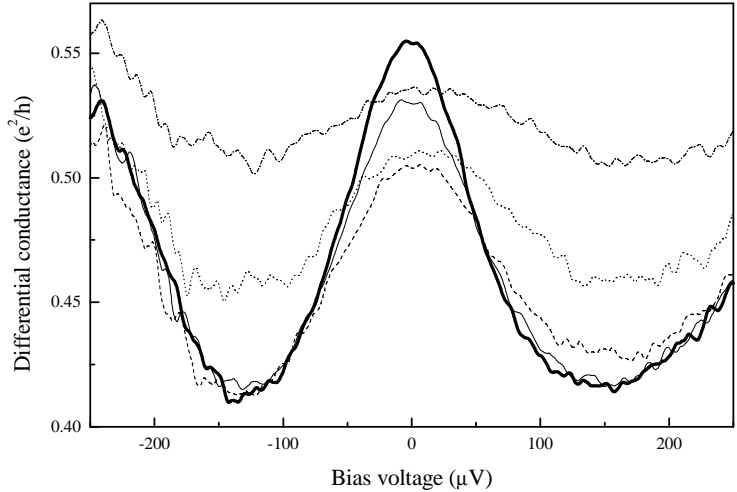,width=16cm,rheight=6.7cm}}
\caption{  
Differential
conductance, $dI/dV_{sd}$, versus bias voltage, $V_{sd}$, for various
temperatures: $T$ (in mK) = 18 (thick), 104 (thin), 154 (dashed), 203
(dotted), 285 (dashed-dotted). All traces are taken in the middle of the
central valley in Fig. 3.}
\end{figure}

The Kondo zero-bias peak in $dI/dV_{sd}$ is shown in Fig. 4 for various
temperatures between 18 mK and 285 mK, and for $V_{g}$ close to the middle
of the central valley in Fig. 3. The increase in $dI/dV_{sd}$ for $\mid
V_{sd}\mid >$150 $\mu $V corresponds to the threshold voltage to overcome
the Coulomb energy $U$. When $T$ is raised, the zero-bias peak decreases and
becomes broader. At the same time, the ``background'' conductance goes up,
due to thermally activated transport. As a result of these competing trends,
the conductance at zero bias already starts to go up before the Kondo peak
is fully suppressed. Even at 285 mK a small bump around zero bias remains
visible.

Figure 5 shows the differential conductance on gray-scale versus both gate
voltage and bias voltage. Because of the rather transparent tunnel barriers,
the usual ``diamond'' shape of the Coulomb blockade regions cannot be
distinguished very well. We estimate that the charging energy $U=e^{2}/C$ $%
\approx $ 500 $\mu $eV, a factor of three smaller than in the weak-tunneling
regime\cite{tjerk2}. The broadening of the single-particle states can be
estimated from the width of the Coulomb blockade (CB) peaks; $h\Gamma
\approx 200$ $\mu $eV. The Kondo zero-bias peak is only present in the
central valley, indicating that this valley corresponds to an odd number of
electrons. The width of the peak yields $k_{B}T_{K}\approx 100$ $\mu $eV, or 
$T_{K}\approx 1$ K. From measurements in the weak tunneling regime we find
an effective electron temperature, $T_{eff}$, of about 70 mK. So, the
thermal energy is $k_{B}T_{eff}\approx 6$ $\mu $eV, i.e. 
much smaller than the other energy scales.

\subsection{Kondo Effect with Microwaves}

The next step is to study the modification of the Kondo effect by microwave
irradiation. We can apply frequencies up to 50 GHz, i.e. the photon energy $%
hf$ can be tuned from 0 to about 200 $\mu $eV. Figure 6a shows the linear
conductance in the Kondo valley, as well as part of the two adjacent
valleys, in the presence of microwaves of frequency 49.5 GHz. For this
frequency, $hf$ $\approx h\Gamma \approx 2k_{B}T_{K}<U/2$, which means that
in the middle of the Kondo valley, the microwaves are not able to ionize the
dot (i.e. $hf<\mu -\varepsilon _{0}\approx U/2$). In this case the dominant
decoherence mechanism due to the microwaves is probably spin-flip
cotunneling.

\begin{figure}[t]
\centerline{\psfig{figure=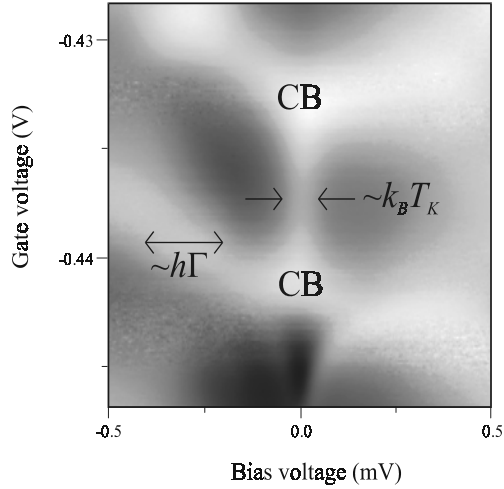,width=11cm,rheight=8.5cm}}
\caption{  
Differential
conductance in gray-scale versus gate voltage and bias voltage. In the
darkest regions, $dI/dV_{sd}\approx 0.35e^{2}/h$, in the lightest regions $%
dI/dV_{sd}\approx 0.7e^{2}/h$. The broad diagonal lines correspond to the
edge of the Coulomb gap. The thin line in the central valley along zero
bias, which is absent in the adjacent valleys, represents the Kondo peak.}
\end{figure}

Different traces correspond to different intensities of the radiation. The
thick curve is taken with an attenuation of -110 dB, which is essentially
the same as without microwaves. If the power is increased, the
two outer valleys go up, as indicated by the arrows, whereas the middle
valley first comes down and only goes up for the highest power. Also, we see
that the two Coulomb peaks decrease and move apart. This behavior
demonstrates that the microwaves suppress the Kondo effect; at a first
glance in a similar way as increasing temperature.

Figure 6b shows a similar set of traces for a lower frequency $f$ = 15.2
GHz. 
Here the photon energy is considerably smaller than $h\Gamma $ and $U$
and of the same order as $k_{B}T_{K}$. 
Nevertheless, we observe virtually
the same behavior as in Fig. 6a for 49.5 GHz, suggesting that the mechanism
for the suppression of the Kondo effect is frequency independent. 

The effect of microwaves on the Kondo peak in the differential conductance
can be seen in Figs. 6c and 6d. The traces are taken in the middle of the
Kondo valley. For increasing microwave power, the Kondo peak decreases and
broadens. We do not observe any sidebands in the differential conductance.
In Fig. 6c these should occur at $V_{sd}=\pm hf/e=\pm 205$ $\mu $V and in
Fig. 6d at $\pm 42$ $\mu $V. Similar results were also obtained for $f=$
19.8 GHz, 31.0 GHz and 41.9 GHz. This set of frequencies was chosen in order
to cover different regimes, starting with $hf$ being the smallest energy
scale up to $hf$ of order $h\Gamma $. The results are the same in all cases.
In none of our measurements we have observed any indication of the
formation of sidebands.

Besides these measurements in the middle of the Kondo valley, we have also
studied the differential conductance more towards the right Coulomb peak. In
this region $hf>\mu -\varepsilon _{0}$ for $f$ = 49.5 GHz, such that
ionization processes by microwaves are, in principle, possible. However, this did not seem to have
a significantly different effect on the behavior of the Kondo peak as a function of microwave power. Also in this
case, no evidence for sidebands of the Kondo resonance  
was found. 

Figure 7a shows in detail the suppression of the Kondo zero-bias peak versus
microwave attenuation. The lower axis gives the value of the applied
attenuation. The upper axis gives the amplitude of the AC signal, $V_{\omega
}$, relative to the amplitude without attenuation, $V_{0}$, on a logarithmic
scale. The exact value of $V_{0}$ cannot be determined, since we do not know
the amplitude of the radiation as seen by the electrons in the dot. This
depends not only on the applied attenuation, but also on the transmission
for the frequency that is used. Although our high-frequency coaxial cable
has a nearly flat frequency response, the coupling of the AC signal from
coaxial cable via the capacitor and gate transmission line to the quantum
dot is probably very frequency dependent\cite{pumping}.

\begin{figure}[p]
\centerline{\psfig{figure=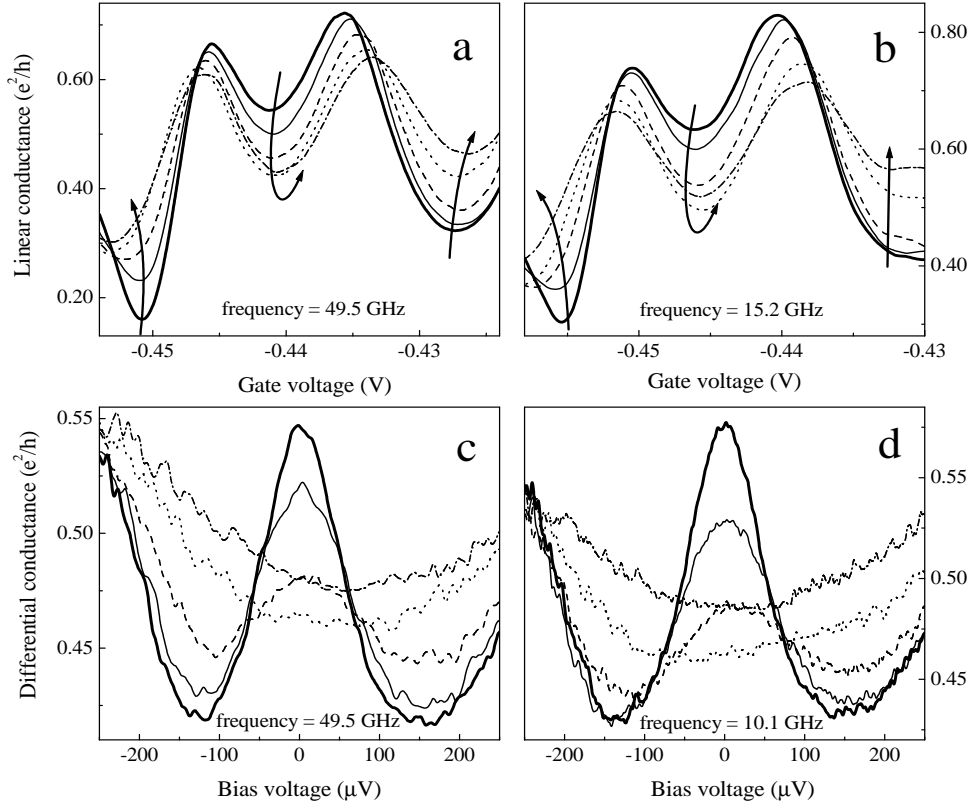,width=19cm,height=26cm,rheight=12.5cm}}
\caption{  
Linear
and differential conductance in the presence of microwaves of several
frequencies. Different curves correspond to different values of the
microwave intensity. Arrows indicate the direction of increasing microwave
power. (a) $f=$49.5 GHz, attenuation (in dB) is equal to: -110 (thick), -20
(thin), -10 (dashed), 0 (dotted), +5 (dashed-dotted). (b) $f=15.2$ GHz,
attenuation (in dB) is: -110 (thick), -50 (thin), -40 (dashed), -30
(dotted), -25 (dashed-dotted). (c) same as in (a). (d) $f=10.1$ GHz,
attenuation (in dB) is: -110 (thick), -56 (thin), -46 (dashed), -36
(dotted), -26 (dashed-dotted). All differential conductance traces are taken
in the middle of the Kondo valley. 
For the highest powers applied, multiple-photon absorption can not be ruled out.}
\end{figure}

We can crudely estimate $V_{0}$ in the following way; the microwaves start
to have an effect when the amplitude of the energy oscillation exceeds the
thermal fluctuations, i.e. when $eV_{\omega }>k_{B}T_{eff}\approx 6$ $\mu $%
eV. From Fig. 7a we find that for $f=19.8$ GHz the onset of the suppression
occurs around $V_{\omega }\approx 3\times 10^{-3}V_{0}$. Equating this to $%
k_{B}T_{eff}/e$ gives $V_{0}\approx 2$ mV. With the same procedure we can
also find $V_{0}$ for the other frequencies. Using these values for $V_{0}$
we obtain absolute values for the amplitude $V_{\omega }$, allowing us to
replot the data of Fig. 7a against the dimensionless parameter $eV_{\omega}/hf$. 
From this procedure we find that curves taken at different
frequencies nearly fall on 
\begin{figure}[t]
\centerline{\psfig{figure=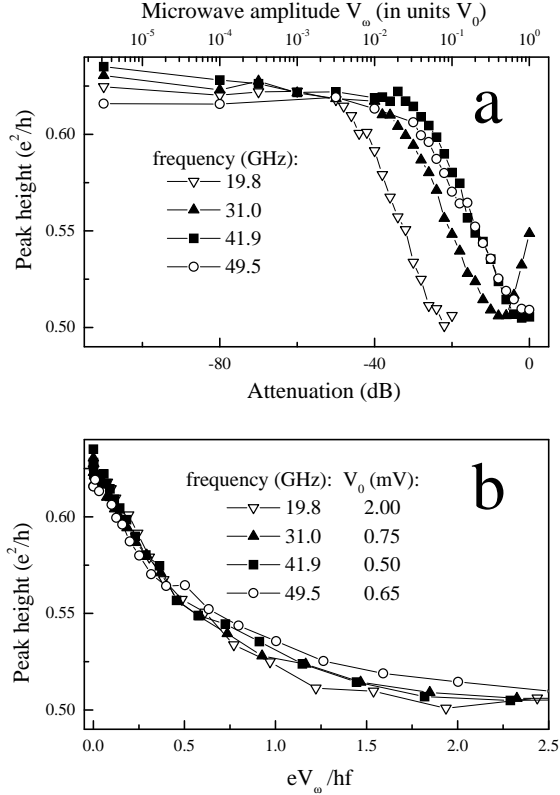,width=10.5cm,rheight=10.3cm}}
\caption{  
(a) Height of the Kondo zero-bias peak versus microwave attenuation
(lower scale). The upper scale gives the amplitude $V_{\omega }$ of the AC
signal in units of the frequency-dependent amplitude $V_{0}$ without
attenuation. (b) Same data, but now plotted versus $eV_{\omega }/hf$. The listed values 
of $V_{0}$ correspond to the onset 
of peak-height suppression in panel (a). 
These parameters were slightly adjusted  
to obtain the best agreement between the different curves.
}
\end{figure}
top of each other. 
To illustrate this mapping
most clearly we have allowed that $V_{0}$ differs somewhat from the value
obtained from the suppression onset in Fig. 7a. 
The results are shown in
Fig. 7b on linear scales. As can be seen, we now obtain a scaling behavior
for the Kondo peak height versus $eV_{\omega }/hf$. We note that this scaling
behavior is not limited to the regime of weak power, i.e. the one-photon
regime where $eV_{\omega }/hf\ll 1$.

We now compare the effect of temperature to the effect of microwaves in more
detail. We note that in Fig. 6a the valley conductance decreases to $%
0.43e^{2}/h$, significantly smaller than $0.49e^{2}/h$, which is the lowest
conductance in the $T$-dependence of Fig. 3. The same difference can also be
seen in the differential conductance plots. In Fig. 4 the zero-bias
conductance starts to rise before the Kondo peak is completely suppressed,
whereas in Fig. 6b the curve with the lowest zero-bias conductance is
essentially flat. This suggests that the Kondo effect is suppressed more
efficiently by microwaves than by temperature. In other words, we cannot
exactly translate microwave power to an ``effective temperature''.
This is to be contrasted with the conclusions of Ref. 15,
where the effect of microwave radiation was indistinguishable from heating.

\begin{figure}[t]
\centerline{\psfig{figure=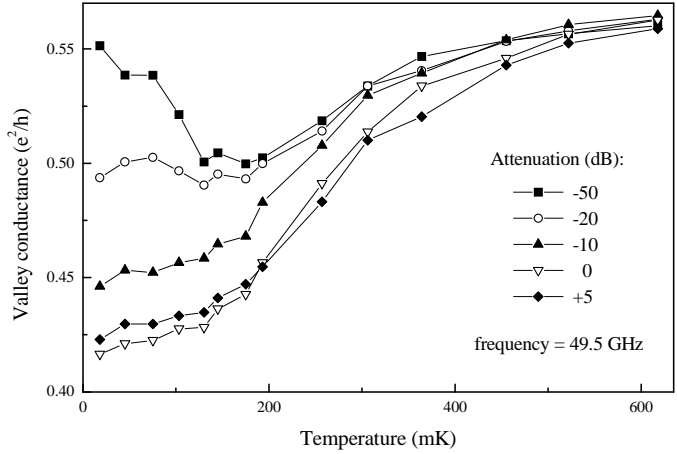,width=14cm,rheight=6.2cm}}
\caption{  
Conductance in the middle of the Kondo valley versus temperature, 
in the presence of microwaves of different intensities.}
\end{figure}

This also follows from Fig. 8, which shows the conductance in the middle of
the Kondo valley versus temperature, in the presence of microwaves of
various intensities. For the lowest intensity (-50 dB attenuation), the
anomalous Kondo temperature-dependence is observed; first the conductance
decreases to about $0.5e^{2}/h$ at 200 mK, and then it starts to increase.
For the curve with -20 dB attenuation, the conductance is already $\sim
0.5e^{2}/h$ at base temperature. It remains more or less constant up to 200
mK, and then follows the -50 dB curve. However, by increasing the microwave
power even more, we can suppress the conductance to a value substantially
lower than $0.5e^{2}/h$. This is not possible by only raising the
temperature, which again demonstrates the difference between heating and
microwaves. At high temperatures, all curves essentially merge together. In
this region, the effect of the microwaves is negligible compared to
temperature.

\section{CONCLUSIONS}

We have studied the modification of the Kondo effect in a quantum dot by
microwave radiation. We have performed measurements over a range of
temperatures, microwave frequencies, and power in order to cover the
different regimes for photon-assisted tunneling processes. In all these
measurements we find no indication for photon-induced sidebands to the Kondo
resonance. The presence of microwaves suppresses the Kondo effect for all
our frequencies. A detailed comparison shows that the microwave-induced
suppression is different from the suppression by higher temperatures. We
find an interesting scaling behavior; the Kondo resonance decreases with the
dimensionless parameter $eV_{\omega }/hf$ independently of frequency.

We believe that the observed suppression is due to a microwave-induced
decoherence of the Kondo resonance by spin-flip cotunneling processes. For a
definite conclusion it is necessary to perform a detailed comparison with
numerical results on the microwave-induced dephasing of the Kondo resonance
at finite temperature. Such a numerical study is presently being performed%
\cite{lopez}.

\begin{center}
{\bf ACKNOWLEDGMENTS}
\end{center}

We thank  Yu. V. Nazarov, L. I. Glazman, A. Kaminski, R.
Aguado, R. L\'{o}pez, and G. Platero for useful discussions. This work was
supported by the Dutch Organization for Research on Matter (FOM), and by the
EU via the TMR network.


\begin{thebibliography}{99}
\bibitem{T-shift}  Each of the traces in Fig. 3 has been given an
appropriate offset in the x-direction, to compensate for a $T$-dependent
shift of all features towards more negative $V_{g}$. The shift is related to
our ``gate-voltage dividers'', consisting of two resistors attached to each
gate (10 M$\Omega $ to ground and 40 M$\Omega $ to the gate voltage source),
thermally anchored to the mixing chamber of the dilution refrigerator. They
provide a low-temperature division of the gate voltage. Due to a $T$%
-dependence of the resistors between roughly 100 mK and 600 mK, the division
factor is not always exactly equal to 5 but varies up to 10\%. Recently we
have checked that without the gate voltage dividers, no $T$-dependent shift
occurs.
\bibitem{pumping}  An extremely frequency-sensitive phenomenon is ``pumping''%
\cite{tjerk}. If the AC voltage drop over the two barriers is slightly
asymmetric, the microwaves lead to a net current even for $V_{sd}=0$; i.e.
the dot acts as an electron pump. Both the value and the polarity of the
pumping current depend very sensitively on $f$. We attribute this to
standing waves in the sample holder. We use frequencies that do not produce
a pumping current. In this case, 
the microwave field does not induce an oscillating source-drain bias voltage.
\bibitem{T-shift2}  The $T$-shift \cite{T-shift} was not present in this
case. Intense microwaves can heat the electron system, but they do not heat
the mixing chamber directly, so they do not affect the gate-voltage dividers.
\end{thebibliography}
\end{document}